\documentclass[letterpaper, 10pt, conference]{ieeeconf}

\IEEEoverridecommandlockouts
\usepackage{graphicx}
\usepackage{hyperref}
\usepackage{amsmath}
\usepackage{amssymb}
\usepackage{xcolor}
\usepackage{comment}
\usepackage{pgfplots}
\usepackage{mathtools}
\pgfplotsset{compat=1.18}

\usepackage{algorithm}
\usepackage{algpseudocode}

\newtheorem{theorem}{Theorem}[section]
\newtheorem{lemma}[theorem]{Lemma}
\newtheorem{proposition}[theorem]{Proposition}

\newtheorem{definition}[theorem]{Definition}
\newtheorem{remark}{Remark}[section]

\newcommand{\real}{\ensuremath{\mathbb{R}}}

\newcommand{\realnonnegative}{\ensuremath{\mathbb{R}}_{\ge 0}}

\newcommand{\until}[1]{\{1,\dots, #1\}}

\newcommand{\supscr}[2]{#1^{\textup{#2}}}
\newcommand{\setdef}[2]{\{#1 \; | \; #2\}}

\newcommand{\dist}{\operatorname{dist}}

\renewcommand{\epsilon}{\varepsilon}

\usepackage{psfrag}

\renewcommand{\P}{\ensuremath{\mathcal{P}(\Omega)}}

\newcommand{\diff}{\ensuremath{\mathrm{d}}}
\newcommand{\supp}{\operatorname{supp}}

\newcommand{\Id}{\operatorname{Id}}
\newcommand{\Int}{\operatorname{Int}}
\renewcommand{\dist}{\operatorname{dist}}
\newcommand{\monge}{\operatorname{(MP)}}
\newcommand{\kl}{\operatorname{KL}}

\newcommand{\kantor}{\operatorname{(KP)}}
\newcommand{\brenier}{\operatorname{(DP)}}

\newcommand{\Pvar}{\ensuremath{\mathcal{P}(\Omega^2)}}
\newcommand{\projone}{\ensuremath{\operatorname{proj}_1}}
\newcommand{\projtwo}{\ensuremath{\operatorname{proj}_2}}
\newcommand{\bulletsym}{\ensuremath{\bullet}}
\newcommand{\bulletend}{%
  \relax
  \ifmmode
  \else
    \unskip\hfill
  \fi
  \bulletsym
}

\newcommand{\oprocendsymbol}{\hbox{$\bullet$}} 
\newcommand{\oprocend}{\relax\ifmmode\else\unskip\hfill\fi\oprocendsymbol}

\makeatother

\title{\LARGE Multiscale distributed MPC for Optimal Transport}

\author{Guillem Pascual, Yu-Jen Huang, Sonia Martínez\thanks{
    G. Pascual and S. Mart{\'\i}nez are with the Department of
    Mechanical and Aerospace Engineering at the University of
    California, San Diego, CA, USA. (e-mail:
    \{gpascualastivill,  soniamd\}@ucsd.edu.),
    Y.-J. Huang is with Oksi (e-mail:
will.huang@oksi.ai) .}}

\begin{document}

\maketitle

\begin{abstract}
  In this paper, we present a novel multiscale formulation for the
  Optimal Transport (OT) of large swarms.  The scheme is divided in
  two levels, a top-level OT formulation with a terminal penalty cost
  to handle time-varying
  densities, 
  and a low-level, Model Predictive Controller (MPC) that allows
  agents to track the nominal velocity under control affine dynamics
  and rejects external disturbances. We show that
   the terminal-penalty dynamic formulation reduces to a regularized
static dual, which agents solve in a distributed way to compute the nominal
velocity.
  We also prove distributional input-to-state stability in the
  $1$-Wasserstein metric of the density flow with respect to the
  mismatch that the low-level and the target motion introduce, so that
  the tracking error only perturbs the macroscopic flow by a bounded,
  quantifiable amount. Finally, we demonstrate via simulations the
  effectiveness of the proposed method handling time-varying targets
  and state perturbations.
\end{abstract}
\section{Introduction}

Composed of a number of relatively simple agents, swarms display rich
collective behavior from local interactions and coordinated
control. However, finding effective mechanisms to shape their
performance remains a challenge, especially in applications that
involve a large number of agents or degrees of freedom. Particularly,
this requires bridging the gap between the control of multitude,
microscopic agent dynamics and that of the macroscopic
population distribution.
Motivated by this,
we propose a multiscale framework for swarm density-tracking problems
while accounting for microscopic control affine dynamics and external
disturbances.

Controller design for large swarms is usually handled via the
mean-field approach~\cite{KE-SB:19, HH-HW:08}.  The resulting density
is then shaped via PDE control~\cite{PF-MK:11}, Markov
Chains~\cite{SB-SJC-FH:17}, Optimal Transport
(OT)~\cite{YC-TTG-MP:21}. 
Mostly, these works abstract away the details of agent-level
  implementations and directly rely on the macroscopic design. In
  contrast, recent approaches use the macroscopic framework to derive
  reduced algorithms that are implementable at the agent level.
  This connection is often established through the use of kernels that
  estimate the density from agent
  locations~\cite{UE-BA:17,RB-GC-GN:25}. A discussion on the
  macroscopic/microscopic relationship can be found
  in~\cite{VK-SM:22}.

  While most of the large swarm control literature focuses on static
  target distributions, recent works have paid attention to
  time-varying distribution tracking.  This
  includes~\cite{ME-JJ-BB:24}, where an MPC is proposed at the
  distributional level through a $2$-Wasserstein running cost. The
  resulting transport is solved in a centralized way to be followed by
  agents, and lacks a characterization of the tracking error.  More
  recently,~\cite{IN-MB:26} relies on the macroscopic OT framework to
  provide a novel coverage control algorithm for time-varying
  densities via a Sinkhorn-like algorithm. The approach makes use of a
  feedforward mechanism and knowledge of density derivatives to
  achieve tracking.  This type of regularization is similar to the one
  employed in~\cite{KI-KK:23} for densities over discrete spaces,
  where MPC is used by LTI
  agents. 
  Both previous works compute the transport plan on the empirical
  measure of the agents, and their guarantees hold for a fixed number
  of agents rather than for the large swarm density. In the context of
  diffusion, other approaches include flow-matching~\cite{KE:25},
  which handles control affine dynamics macroscopically via the
  Liouville control equation, then learns its feedback offline from
  trajectories that realize a prescribed interpolation between two
  measures. In contrast, our method relies only on the current target
  snapshot within each agent's neighborhood, without its velocity, and
  bounds the tracking error of the swarm density, including the lag
  caused by the target motion, in the $1$-Wasserstein metric.

  The design of macroscopic algorithms usually relies on the OT
  framework~\cite{CV:08}.  The dynamic formulation of
  OT~\cite{JDB-YB:00} becomes especially relevant for density tracking
  applications; the inclusion of a penalty term as in Model Predictive
  Control results into a type of proximal OT problem as
  in~\cite{JDB:03} (for $L^2$ penalties) or as
  in~\cite{JB-PD-MAK-YP-LRB:22,RB-PB-MAK-LRB-BJZ:25} (for divergence
  penalties) from the generative learning literature.  The robustness
  of these algorithms to disturbances can now be quantified in terms
  of a novel distributional Input-to-State stability~\cite{GP-SM:26},
  also related to incremental stability~\cite{YK-SB-AD-FB:26}, which
  we exploit here.

In this work, we consider a density tracking problem in terms of the
$W_1$ metric, which, unlike previous work, quantifies the cost in terms of
distance traveled.
In order to track a moving target, we add a
terminal penalty to the dynamic OT formulation, in a similar way as done for MPC.
By choosing adequately this penalty, we show the problem is
equivalent to a regularized convex OT dual problem, which allows for
the distributed computation of the transport velocity when
approximated by compactly supported kernels.  This is enlarged with an
MPC at the agent level, for finite-horizon velocity tracking, which
can handle control affine dynamics as well as external disturbances.
After this, we characterize the effects of the disturbances that the
low-level MPC introduces at the macroscopic level. In particular, we
perform a distributional input-to-state stability analysis in
Wasserstein space that guarantees a bounded tracking error for
bounded-velocity targets.  Simulations show the scalability of the
design to large-scale multi-agent systems, in a fast and flexible way
exhibiting a robust behavior to time-varying targets and external
disturbances.

\section{Notation}
We let $\real$, $\realnonnegative$ and $\real^d$ denote the real
numbers, nonnegative reals, and the $d$-dimensional Euclidean space,
with $\dist(x,y)$ the Euclidean distance.  The gradient and divergence
operators are denoted as $\nabla$ and $\nabla\cdot$, and
$\partial_t:=\partial/\partial t$.  We denote the set of probability
measures over a convex compact domain $\Omega$ as
$\mathcal{P}(\Omega)$; note that in a compact domain $n$-th moments
are bounded. Denote the interior, closure, and boundary of a set
$A \subseteq \real^d$ as $\Int{(A)}$, $\overline{A}$, and $\partial A$;
respectively, and the volume of the set is denoted $|A|$.  Denote the
support of a function as
$\supp(\rho) = \overline{\{ x \mid \rho(x) > 0 \}}$. The Lebesgue
measure on $\real^d$ is denoted by $\vartheta\equiv \diff x$, and 
$\diff \rho$ (or, with a slight abuse of notation,
 simply $\rho$) refers to the measure with density function $\rho$ when
$\diff\rho \ll \vartheta$; that is, when $\diff\rho$ is absolutely
continuous with respect to the Lebesgue measure.  The pushforward of
a measure $\mu$ towards $\nu$ under a measurable
$T:\Omega \to \Omega$ is denoted $T_{\#}\mu=\nu$; that is, for any
measurable subset $B$ of $\Omega$, $\nu(B)=\mu(T^{-1}(B))$, and $\Id$
is the identity map. In
particular, denoting by $\projone(x,y)=x$ and $\projtwo(x,y)=y$ the
canonical projections of $\Omega\times\Omega$ onto $\Omega$, the
marginals of $\pi\in\Pvar$ are $(\projone)_{\#}\pi$ and
$(\projtwo)_{\#}\pi$.  For $p \geq 1$, the space
$L^p(\Omega; \diff\rho)$, or simply $L^p(\rho)$, is the space of
measurable functions $f$ such that
$\|f\|_{L^p(\rho)} := (\int_\Omega |f|^p \diff\rho)^{1/p}<\infty$,
while $L^p(\Omega)$ is with respect to $\vartheta$ and
$\|\cdot\|_\infty$ is the essential supremum norm.  The Sobolev space
is denoted $W^{1,\infty}(\Omega)$, formed by functions
$f\in L^\infty(\Omega)$ whose weak first derivatives belong to
$L^\infty(\Omega)$.  Given a functional
$F: \mathcal{P}(\Omega)\to\real$, its first variation is denoted as
$\frac{\delta F}{\delta\rho}$.
Finally, a continuous function
$\psi:\mathbb{R}_{\geq0}\to\mathbb{R}_{\geq0}$ is of class
$\mathcal{K}$ if it is strictly increasing with $\psi(0)=0$, and of
class $\mathcal{K}_\infty$ if in addition it is unbounded.

\section{Preliminaries}
In this section, we present the main results in Optimal Transport that
will be used in this work. Further details, such as functionals over
the distributional space and its differentiability, can be found
in~\cite{GP-SM:26}.  For the proofs and further explanation of the
results here,  see~\cite{FS:15}.
\begin{definition}[Transport Map]
  Let $ \rho_{0}, \rho_{1}\in \P$. A measurable map
  $T:\Omega\to\Omega$ is a
  transport map from $\rho_0$ to $\rho_1$ if
  $T_{\#}\rho_{0}=\rho_{1}$. \oprocend
\end{definition}
A transport map is optimal if it minimizes  certain cost. The following
is the original Monge formulation of the Optimal Transport problem.
\begin{definition}[Monge Problem]
  \label{def:Monge}
  Given two probability measures
  $\rho_{0}, \rho_{1} \in \P$ and a cost
  function $c:\Omega\times\Omega\to\real_{\ge0}$,
  the Monge problem is defined as
\begin{equation}\label{eq:OT}
  \begin{aligned}
  \monge \quad\inf_{T} \quad & \int_{\Omega} c\bigl(x,T(x)\bigr) \, \diff\rho_{0}(x) \\
  \text{subject to} \quad & T_{\#}\rho_{0}=\rho_{1}.
  \end{aligned}
\end{equation}
When a solution to $\monge$ exists, it is called an \textit{optimal
transport map}, $T_*$.\oprocend
\end{definition}
The transport constraint is nonlinear in $T$, the admissible set is in
general nonconvex and it may be empty.  Even when admissible maps
exist, an optimal one need not.  The Kantorovich formulation relaxes
the problem to couplings.
\begin{definition}[Coupling]
  Let $\rho_{0}, \rho_{1}$ be probability distributions in
  $\P$. A \textit{coupling} $\pi\in\Pvar$ is a probability distribution such that
  $(\projone)_{\#}\pi=\rho_0$ and
  $(\projtwo)_{\#}\pi=\rho_1$.
  We denote the set of couplings of $\rho_0,\rho_1$
  as $\Pi(\rho_0,\rho_1)$.\oprocend
\end{definition}

\begin{definition}[Kantorovich Problem]
  Given two probability distributions $\rho_{0},\rho_{1}\in\P$,
  and a cost function $ c:\Omega\times\Omega\to\real_{\ge0}$,
  the Kantorovich problem is:
  \begin{equation}\label{eq:kOT}
    \kantor\quad\inf_{\pi\in\Pi(\rho_0,\rho_1)} \quad \int_{\Omega^2} c\bigl(x,y\bigr) \, \diff\pi(x,y).
  \end{equation}
  The solution is denoted $\pi^*$.  Note that this relaxation yields a
  linear program over a set that is always nonempty and convex.
  Furthermore, this relaxation admits a dual formulation, that
  satisfies strong duality:
  \begin{equation}
  \label{eq:Kantorovich-dual}
  \begin{aligned}
  \kantor^* \quad\sup_{\phi,\psi\in L^1(\Omega)}
  \quad &
  \int_{\Omega}\phi(x)\,\diff\rho_0(x)+\int_{\Omega}\psi(y)\,\diff\rho_1(y) \\
   \text{subject to}\quad&\phi(x)+\psi(y)\le c(x,y),
  \end{aligned}
  \end{equation}
  where $\phi$ is the \textit{Kantorovich potential} between $\rho_0$
  and $\rho_1$ and $\psi=\phi^c$ is its \textit{$c$-conjugate} pair,
  which satisfies:
  \begin{equation*}
	\phi^c(y) := \inf_{x\in\Omega}\big\{c(x,y)-\phi(x)\big\}. \tag*{\oprocend}
  \end{equation*}
\end{definition}

A simplification of the dual problem exists for the case
$c(x,y)=|x-y|$, as the next result states, see~\cite{VK-SM:25}:
\begin{lemma}[Lipschitz Bound]
\label{lem:local-lipschitz}
Let $c(x,y)=|x-y|$. 
It holds that $\phi^{c}=-\phi$, and
\begin{equation*}
|\phi(x)-\phi(y)|\leq |x-y|,
\qquad \forall x,y\in\Omega.
\end{equation*}
Furthermore, the Kantorovich potential is differentiable almost
everywhere in the interior $\Int{(\Omega)}$, with
\begin{equation*}
|\nabla\phi(x)|\leq 1, \qquad \text{a.e. in }\Int{(\Omega)}. \tag*{\oprocend}
\end{equation*} 
\end{lemma}
Using the previous lemma, we can reformulate the dual problem for this
particular cost as follows:
\begin{equation}
\label{eq:dual-rn}
\begin{aligned}
  C(\rho_0,\rho_1)=\sup_{\phi \in W^{1,\infty}(\Omega)}\quad &\int_{\Omega}\phi\left(\diff\rho_0-\diff\rho_1\right)\\
  \text{subject to} \quad &
                            |\nabla \phi| \leq 1,\qquad \text{a.e. in } \Omega.
\end{aligned}
\end{equation}
This formulation of the Optimal Transport problem is convenient for
distributed implementation, as we discuss later.  Furthermore, the
problem also admits a dynamic formulation as follows. Pick any $T>0$,
let $c(x,y)=|x-y|$, and define:
\begin{equation}
  \label{eq:dynamic-w1}
  \begin{aligned}
    \brenier \quad \inf_{\rho, v} \quad & \int_0^T \int_{\Omega} |v(t,x)| \diff\rho(t,x) \diff t \\
    \text{s.t.} \quad & \partial_t \rho(t,x) + \nabla \cdot \bigl( \rho(t,x) v(t,x) \bigr) = 0, \\
                                   & \rho(0,x) = \rho_0, \qquad \rho(T,x) = \rho_1,
  \end{aligned}
\end{equation}
where $\rho(t,x)$ represents the density at time $t$, and $v(t,x)$ is
its velocity field. Note that under the change of variables
  $m = \rho v$, $\brenier$ is a convex optimization. 
  Finally, note that under the assumptions of absolute continuity of
  $\rho_0,\rho_1$, the compactness and convexity of $\Omega$, and the
  choice of $c(x,y) = |x-y|$, the three formulations~\eqref{eq:OT},
  \eqref{eq:kOT}, and~\eqref{eq:dynamic-w1} admit solutions,
and their optimal values correspond to
$W_1(\rho_0,\rho_1)$~\cite{FS:15}, the $1$-Wasserstein distance.

\section{Problem formulation}
We consider a swarm of $N$ agents with state $z_i$,
$i\in\until{N}$, and control affine dynamics:
\begin{equation*}
	\dot{z}_i=f(z_i)+g(z_i)u_i.
\end{equation*}
The swarm is described by a density $\rho$, which aims to track a
moving target $\rho^*$ evolving under a continuity equation:
\begin{equation}\label{eq:target-traj}
  \partial_t\rho^* =-\nabla\cdot(\rho^*v^*(t,x)),
\end{equation}
for some velocity field $v^*(t,x)$ defined over $\Omega$, and subject
to no-flux boundary conditions.  To achieve this, our first goal is to
design a macroscopic potential field $\phi$ whose negative gradient
$-\nabla \phi$ steers the swarm distribution towards the moving
target. Further, we aim to derive a (fast-time scale) distributed
algorithm to approximate $-\nabla \phi$ that the agents can use to
calculate this nominal velocity. This will be complemented with an
agent-level MPC for velocity tracking subject to control affine
dynamics, and small control effort.

The interconnection of the macroscopic dynamics with the microscopic
algorithms, as well as other external disturbances, will result into a
macroscopic equation:
\begin{equation*}
  \partial_t\rho = -\nabla\cdot(\rho(-\nabla\phi+v(t,x))),
\end{equation*}
subject to no-flux boundary conditions.

Therefore, to ensure the efficacy of the method, and a consistent
design across scales, we aim to guarantee the bounded stability
properties of the algorithm as:
\begin{equation*}
  \lim\sup_{t\to\infty} W_1(\rho,\rho^*)\leq
  \varphi (\|v \|^2_{L^2}+ \|v^*\|^2_{L^2}),\quad\varphi\in\mathcal{K}_\infty,
\end{equation*}
where $W_1$ is the $1$-Wasserstein metric.

To summarize, the combined algorithm design follows a two-scale or
two-level method.  At the macroscopic, top level a
potential field $\phi$ 
guides the swarm distribution toward the target (over a slow time
scale), accounting for the target motion via a terminal penalty. The
potential field computation is carried out on a fast-time scale via a
distributed microscopic algorithm. Further, to account for finite
actuation, dynamics, and perturbations, agents solve for
an individual finite-horizon MPC problem independently.

\section{Multiscale Control for Optimal Transport}
In this section, we discuss the analytical formulation of the
multiscale scheme, and its subsequent discretization that
lends itself to distributed microscopic implementations.
\subsection{High-level OT planner}

Consider a moving density target $\rho^*(t,x)$ that evolves according
to~\eqref{eq:target-traj}. Fix $T_H>0$, and denote the value of the
target and swarm densities at time $T_H$ as $\rho_{T_H}^*$ and
$\rho_{T_H}$, respectively.  In order to account for the moving
target, we add a terminal cost to the dynamic OT
formulation~\eqref{eq:dynamic-w1}, rendering the formulation similar
to an MPC. Let $\mathcal{F}:\P \times \P \rightarrow \real$ be a
proper, convex, 
 and lower semi-continuous functional, define
$ \mathcal{F}(\rho_{T_H}) = \mathcal{F}(\rho_{T_H},\rho^*_{T_H}) $,
for any $T_H>0$. Then, we have:

{\small
\begin{equation}
\label{eq:top-mpc}
\begin{aligned}
  \inf_{\rho, v} \quad & \int_\tau^{\tau+{T_H}} \int_{\Omega} |v(t,x)| \diff\rho(t,x) \diff t+\mathcal{F}(\rho_{\tau+T_H})\\
  \text{s.t.} \quad & \partial_t \rho(t,x) + \nabla \cdot \bigl( \rho(t,x) v(t,x) \bigr) = 0, \\
                       & \rho(\tau,x) = \rho_0.
\end{aligned}
\end{equation}
} 

In what follows, assume $\tau = 0$ for simplicity of notation. The
next proposition states the equivalence between \eqref{eq:top-mpc} and
its dual formulation. Seen through the lens of its dual,
  \eqref{eq:top-mpc} can be interpreted as a proximal optimization
  w.r.t.~$W_1$. In this regard,~\cite{JB-PD-MAK-YP-LRB:22} studies the
  problem equivalence for (small) penalties given by
  Lipschitz-regularized $f$-divergences.  Here, we provide an
  alternative equivalence proof for general convex penalties,
  including the $L^2$ cost, which is not an $f$-divergence. These
  derivations are the starting point of Theorem~\ref{prop:stab}.

	\begin{proposition}\label{prop:fenchel-dual}
	  Let $\mathcal{F}$ be as in~\eqref{eq:top-mpc}, and define the
	  (convex) functional $\mathcal{G}:W^{1,\infty}(\Omega)\rightarrow\real$ as:
	  \begin{equation}	\label{eq:fenchel_final_relation}
		\mathcal{G}(\phi) = \sup_{\rho\in\P} \left\{ \int_{\Omega}
	          \phi(x) \bigl( \rho(x) - \rho_{T_H}^*(x) \bigr) \diff x
	          - \mathcal{F}(\rho) \right\}.
	      \end{equation}
	      Assume that, for every $\phi\in W^{1,\infty}(\Omega)$, the
	      supremum in~\eqref{eq:fenchel_final_relation} is attained at a
	      unique $\sigma\in\P$ a.e.~and $\mathcal{G}$ is differentiable
	      with well-defined first variation
	      $\frac{\delta\mathcal{G}}{\delta\phi}(\phi)=\sigma-\rho^*$, see~\cite{JFB-AS:13}.
	      \footnote{Note that
	        $\mathcal{F}$ is convex and the first integral is linear in
	        $\rho$, the uniqueness of $\sigma$ should hold under
	        sufficient regularity and strict convexity of
	        $\mathcal{F}$. Similarly, the differentiability of
	        $\mathcal{G}$ follows from Danskin's theorem.} Define the
	      regularized static Kantorovich dual~\eqref{eq:Kantorovich-dual},
	      as follows:
	    \begin{equation}
	    \label{eq:dual-mpc-general}
	    \begin{aligned}
	      \sup_{\phi \in W^{1,\infty}(\Omega)}\quad
	      J(\phi) :=&\int_{\Omega}\phi\left(\diff\rho_0-\diff\rho^*_{T_H}\right)
	        -\mathcal{G}(\phi),\\
	      \text{subject to} \quad &
	                        |\nabla \phi| \leq 1,\qquad\text{a.e. in } \Omega.
	    \end{aligned}
	  \end{equation}
  Then, both optimization problems~\eqref{eq:dual-mpc-general}
      and~\eqref{eq:top-mpc} are equivalent, that is, the solution
      exists and it attains the same value.
 \end{proposition}
   
\begin{proof}
  For simplicity, consider the feasible set:
\begin{equation*}
  K:=
  \setdef{\phi\in W^{1,\infty}(\Omega)}{ |\nabla\phi|\le 1 \text{ a.e. in }
    \Omega}.
\end{equation*}
Note that for a convex set $\Omega$ the set $K$ corresponds to
$1$-Lipschitz functions.
We first show that~\eqref{eq:dual-mpc-general} admits a maximizer.
Existence relies on the invariance of the dual problem under
constants, which is given directly by the definition of $\mathcal{G}$:
for any $c\in\real$ and $\rho\in\P$,
$\int_\Omega c\,(\rho-\rho_{T_H}^*)\diff x=0$, since both are
probability measures, and thus $\mathcal{G}(\phi+c)=\mathcal{G}(\phi)$
by~\eqref{eq:fenchel_final_relation}. As
$\int_\Omega c\,(\rho_0-\rho_{T_H}^*)\diff x=0$, as well, then
$J(\phi+c)=J(\phi)$, and we can restrict $ K$ to those $\phi$ with
$\phi(x_0)=0$, for a fixed $x_0\in\Omega$.  It can be shown that these
functions are $1$-Lipschitz (implying continuity) and bounded by
$\operatorname{diam}(\Omega)$, i.e., equicontinuous and uniformly
bounded. The rest follows exactly as in~\cite[Chapter 1]{FS:15} by the
Arzel\`a--Ascoli theorem.

Next, by the definition of $\mathcal{G}$
in~\eqref{eq:fenchel_final_relation}, for any $\rho\in\P$ and
$\phi\in K$ we have the following inequality:
\begin{equation*}
  \mathcal{F}(\rho)\geq \int_\Omega\phi(\rho-\rho_{T_H}^*)
  \diff x - \mathcal{G}(\phi).
\end{equation*}
Let $(\rho,v)$ be admissible for~\eqref{eq:top-mpc} and $\phi\in
K$. By the continuity equation, the no-flux condition, and
$|\nabla\phi|\le1$, we have
$\frac{\diff}{\diff t}\int_\Omega\phi\,\rho\diff
x=\int_\Omega\nabla\phi\cdot v\,\rho\diff x
\ge-\int_\Omega|v|\,\rho\diff x$. Integrating over $[0,T_H]$, we obtain:
\begin{equation*}
  \int_0^{T_H}\!\!\int_\Omega|v|\,
  \rho\diff x\diff t\ge\int_\Omega\phi\,(\rho_0-\rho_{T_H})\diff x.
\end{equation*}
Adding the previous two inequalities, the terms in $\rho$
cancel out, so the cost of $(\rho,v)$ in~\eqref{eq:top-mpc} is at
least $J(\phi)$. Taking the infimum over $(\rho,v)$ and the supremum
over $\phi\in K$
yields~\eqref{eq:top-mpc}~$\ge$~\eqref{eq:dual-mpc-general}.

Since $J$ is concave, differentiable, and $K$ is convex, from
the first-order concavity condition and optimality of $\phi^*$:
\begin{equation*}
  \int_\Omega(\psi-\phi^*) \frac{\delta J}{ \delta \phi }\diff x \leq 0, \qquad \forall\,\psi\in K.
\end{equation*}
Since $\frac{\delta J}{ \delta \phi }_{\big| \phi^*} = \rho_0 -
  (\rho_{T_H}^* + \frac{\delta \mathcal{G}}{\delta\phi}_{\big| \phi^*}) = \rho_0 - \sigma$,
we have:

{\small
\begin{equation*}
\int_\Omega(\psi-\phi^*)\big(\rho_0-\rho_{T_H}^*-\frac{\delta \mathcal{G}}{\delta\phi}_{\big| \phi^*}\big)\diff x
=\int_\Omega(\psi-\phi^*)(\rho_0-\sigma)\diff x\le 0.
\end{equation*}
} By construction of $\mathcal{G}$, $\sigma\in\P$; in
particular, $\sigma\ge0$ and $\int_\Omega\sigma\diff x=1$.
The same inequality states
that $\phi^*$ attains the supremum of
$\int_\Omega\psi\,(\rho_0-\sigma)\diff x$ over $\psi\in K$, which,
as $\sigma\in\P$, equals $W_1(\rho_0,\sigma)$
by~\eqref{eq:dual-rn}. Due to this, 
$\phi^*$ is a Kantorovich potential between $\rho_0$ and $\sigma$, and
$W_1(\rho_0,\sigma)=\int_\Omega\phi^*(\rho_0-\sigma)\diff x$.

As $\sigma$ attains the supremum
in~\eqref{eq:fenchel_final_relation} for $\phi=\phi^*$, the equality
$\mathcal{F}(\sigma)=\int_\Omega\phi^*(\sigma-\rho_{T_H}^*)\diff
x-\mathcal{G}(\phi^*)$ holds. 
Adding both yields:
\begin{equation}\label{eq:ineq-1}
W_1(\rho_0,\sigma)+\mathcal{F}(\sigma)  =\int_\Omega\phi^*(\diff\rho_0-\diff\rho_{T_H}^*)-\mathcal{G}(\phi^*).
\end{equation}

 Finally, from the dynamic formulation of
$W_1$~\eqref{eq:dynamic-w1}, 
the running cost of the admissible paths from $\rho_0$ to $\sigma$ has
as infimum $W_1(\rho_0,\sigma)$, that is:
\begin{equation*}
\label{eq:benamou-brenier}
\begin{aligned}
 W_1(\rho_0, \sigma)=  \inf_{\rho, v} \quad &\left\{ \int_0^{T_H} \int_{\Omega} |v(t,x)| \diff\rho(t,x) \diff t \right\} \quad \\
  \text{s.t.} \quad & \partial_t \rho + \nabla \cdot ( \rho v ) = 0, \\
                       & \rho(0,x) = \rho_0, \quad \rho(T_H,x) = \sigma,
\end{aligned}
\end{equation*}
see~\cite[Chapter 4]{FS:15}. Setting $\rho_{T_H}=\sigma$ in~\eqref{eq:top-mpc}, yields:
\begin{equation*}
\eqref{eq:top-mpc}\le W_1(\rho_0,\sigma)+\mathcal{F}(\sigma),
\end{equation*}
which together with~\eqref{eq:ineq-1} yields the inequality $\eqref{eq:top-mpc}\leq\eqref{eq:dual-mpc-general}$,
 the reverse inequality, proving the claim.
\end{proof}

\begin{remark}[Regularization]
  As customary in standard MPC, the introduction of a penalty term is
  motivated by increased stability and enhanced tracking. As we show
  later, its presence becomes critical for the good behavior of the
  numerical solutions. \oprocend
\end{remark}
\subsection{Microscopic and distributed implementation}\label{sec:discr-stab}

Here, we present two types of regularizations; given by an $L^2$ and a
$\kl$ penalty. We start with the following result:
\begin{lemma}\label{le:reg-penalty}
Let $\rho^*=\rho^*_{T_H}$, and $Z:=\int_\Omega e^{\epsilon\phi}\rho^*\diff x$.
 Then $\mathcal{G}$ in~\eqref{eq:fenchel_final_relation} can be expressed as
\begin{equation*}
\begin{aligned}
\mathcal{F}_{L^2}(\rho):={}&
\tfrac{1}{2\epsilon}\|\rho-\rho^*\|_{L^2}^2
\Rightarrow
\mathcal{G}_{L^2}(\phi)
=\tfrac{\epsilon}{2}\|\phi\|_{L^2}^2
\quad\text{if }\widetilde{\sigma}\geq0,\\
\mathcal{F}_{\kl}(\rho):={}&
\tfrac{1}{\epsilon}\kl(\rho\,\|\,\rho^*)
\Rightarrow
\mathcal{G}_{\kl}(\phi)\\
&\hspace{1.5cm}
=\tfrac{1}{\epsilon}\int_\Omega\rho^*
\bigl(e^{\epsilon\phi}-1-\epsilon\phi\bigr)\diff x.
\end{aligned}
\end{equation*}
where $\widetilde{\sigma}=\rho^*+\epsilon\phi$ and 
$\sigma=e^{\epsilon\phi}\rho^*$; respectively, are the
unique maximizer in~\eqref{eq:fenchel_final_relation}, and
$\kl(\rho\,\|\,\rho^*)=+\infty$ if $\rho\not\ll\rho^*$.
\end{lemma}
\begin{proof}
  Define the relation in~\eqref{eq:fenchel_final_relation} as
  $\sup_{\rho\in\P} \chi(\rho)$.  As seen in
  Proposition~\ref{prop:fenchel-dual}, $\mathcal{G}(\phi)$ is
  invariant under constant shifts, so we may normalize $\phi$ as
  desired.  In both cases, $\mathcal{F}$ is strictly convex, so $\chi$
  is strictly concave, and any $\sigma\in\P$ satisfying the
  stationarity condition $\delta\chi/\delta\rho|_\sigma=\lambda$ a.e.,
  with $\lambda\in\real$ the multiplier of $\int_\Omega\rho\diff x=1$
  (see~\cite[Chapter 7]{FS:15}), is its unique maximizer over $\P$,
  since
  $\chi(\rho)\le\chi(\sigma)+\int_\Omega \lambda(\rho-\sigma)\diff
  x=\chi(\sigma)$ for all $\rho\in\P$.
  
  \emph{$L^2$ case:}
  $\delta\chi/\delta\rho|_\sigma=\phi-(\widetilde{\sigma}-\rho^*)/\epsilon=\lambda$
  gives $\widetilde{\sigma}=\rho^*+\epsilon(\phi-\lambda)$, and
  $\int_\Omega\widetilde{\sigma}\diff x=1$, normalizing
  $\int_\Omega\phi\diff x=0$ gives $\lambda=0$. Hence
  $\widetilde{\sigma}=\rho^*+\epsilon\phi\in\P$ iff
  $\widetilde{\sigma}\ge0$, in which case
  $\mathcal{G}(\phi)=\chi(\widetilde{\sigma})=\epsilon\|\phi\|_{L^2}^2-\frac{\epsilon}{2}\|\phi\|_{L^2}^2=\frac{\epsilon}{2}\|\phi\|_{L^2}^2$.

 \emph{KL case:} $\delta\chi/\delta\rho|_\sigma=\phi-\frac{1}{\epsilon}\bigl(\log(\sigma/\rho^*)+1\bigr)=\lambda$
gives $\sigma\propto e^{\epsilon\phi}\rho^*$, and normalization with $Z=1$ gives
$\sigma=e^{\epsilon\phi}\rho^*\in\P$. Using $\log(\sigma/\rho^*)=\epsilon\phi$,
$\mathcal{G}(\phi)=\chi(\sigma)=\int_\Omega\phi(\sigma-\rho^*)\diff x-\int_\Omega\sigma\phi\diff x
=-\int_\Omega\phi\,\rho^*\diff x=\frac{1}{\epsilon}\int_\Omega\rho^*\bigl(e^{\epsilon\phi}-1-\epsilon\phi\bigr)\diff x$,
as $\int_\Omega(e^{\epsilon\phi}-1)\rho^*\diff x=Z-1=0$.
\end{proof}
 For the $L^2$ regularization case, we rely on the assumption
  $\tilde\sigma\geq0$, which the next result characterizes.
\begin{lemma}\label{le:positivity}
Consider the optimization in~\eqref{eq:dual-mpc-general} with $\mathcal{G}_{L^2}$ from Lemma~\ref{le:reg-penalty}.
Then, for a.e.\ $x\in\Omega$,

{\small
\begin{equation*}
  \widetilde{\sigma}(x)=\rho^*(x)+\epsilon\phi^*(x)\geq
  \operatorname*{ess\,inf}_{y\in\Omega}\big\{\rho(y)+\rho^*(x)-\rho^*(y)+\epsilon|x-y|\big\}.
\end{equation*}
} In particular, if $\rho^*$ is $\epsilon$-Lipschitz a.e., then
$\widetilde{\sigma}\geq\operatorname*{ess\,inf}_\Omega\rho\geq0$
a.e. Hence, $\widetilde{\sigma}\in\P$ and $\widetilde{\sigma}$ is the
unique maximizer in~\eqref{eq:fenchel_final_relation} and
$\mathcal{G}_{L^2}(\phi^*)=\frac{\epsilon}{2}\|\phi^*\|_{L^2}^2$.
\end{lemma}

\begin{proof}
  Let $f:=(\rho-\rho^*)/\epsilon\in L^2(\Omega)$. Completing the
  square,
  $J(\phi)=\frac{\epsilon}{2}\|f\|_{L^2}^2-\frac{\epsilon}{2}\|\phi-f\|_{L^2}^2$,
  and $\phi^*$ (the maximizer of~\eqref{eq:dual-mpc-general}) is a point of $K$ closest to $f$ in $L^2$.
  Let $g(x):=\operatorname*{ess\,inf}_{y\in\Omega}\{f(y)+|x-y|\}$; it is
  not hard to see that $g$ is the largest $1$-Lipschitz minorant of
  $f$. 
  Since
  $f\geq-\|\rho^*\|_\infty/\epsilon$, then $g$ is finite  and
  so $g\in K$.  In particular, $\psi:=\max\{\phi^*,g\}\in K$ and
  $\psi=\phi^*$ outside $A:=\{\phi^*<g\}$, while on $A$ we have
  $\phi^*<g\leq f$ a.e. Hence,
  $|\psi-f|=f-g<f-\phi^*=|\phi^*-f|$, and
  $\|\psi-f\|_{L^2}<\|\phi^*-f\|_{L^2}$ unless $|A|=0$. As $\phi^*$ is
  the optimizer, $\phi^*\geq g$ a.e.; multiplying by $\epsilon$ and
  adding $\rho^*$ gives the stated bound.  If $\rho^*$ is
  $\epsilon$-Lipschitz, then $\rho^*(x)-\rho^*(y)+\epsilon|x-y|\geq0$,
  so
  $\widetilde{\sigma}\geq\operatorname*{ess\,inf}_\Omega\rho\geq0$. Together
  with
  $\int_\Omega\widetilde{\sigma}\diff
  x=1+\epsilon\int_\Omega\phi^*\diff x=1$, this gives
  $\widetilde{\sigma}\in\P$, and Lemma~\ref{le:reg-penalty} applies.
\end{proof}

The implementation of the infinite-dimensional problem by the
multi-agent system will use a discretization given by a kernel
approximation~\cite{UE-BA:17}.  Let
$K_h : \real^d \to \realnonnegative$ be a smooth, radially symmetric
kernel with bandwidth $h$ and compactly supported on a ball of
 radius $R$; that is, $\supp (K_h)=\bar{B}_R(0)$. 
Then, we define the approximated density and potential field as follows: 

  {\small
\begin{equation}\label{eq:kernel-funs}
  \rho_N(x) = \frac{1}{N}\sum_{i=1}^N K^a_h(x - x_i),\quad
  \phi_N(x) = \sum_{j=1}^N \alpha_j K^f_h(x - x_j),
  \end{equation}
}where $(\alpha_1, \dots, \alpha_N) \in \real^N$.  Define the
$2R$-disk proximity graph $\mathcal{G}_{2R}(x)=(V,E_{2R}(x))$, where
$V = \until{N}$, and
$E_{2R}(x)=\setdef{(i,j)\in V\times V}{|x_i-x_j|\leq 2R}$, and assume that
two agents can communicate if they are within radius $2R$.  In what
follows, we assume that the kernel basis satisfies convergence of the
density estimation to the mean-field limit (with respect to $W_1$) and that the basis is dense
in $C^1(\Omega)$, for a given~$h$. 

This continuous parameterization allows for a distributed
implementation, as presented in the following.
\begin{lemma}
  Let $J_N(\phi) := \int_\Omega \phi (\diff \rho_N - \diff \rho_{T_H}^*
    )- \frac{\epsilon}{2}\|\phi\|_{L^2}^2$.  The substitution of the
  approximations~\eqref{eq:kernel-funs} into the
  optimization~\eqref{eq:dual-mpc-general} 
  results into the reduced convex problem
  w.r.t.~$\alpha =(\alpha_1,\dots,\alpha_N)$:
\begin{align}
    \label{eq:smooth-kantorovich}
      \max_{(\alpha_1,\ldots,\alpha_N)}\quad
      &J_N(\phi_N) := \sum_{i=1}^{N} \alpha_i
                                                w_i-\frac{\epsilon}{2}
                                               \sum_{i=1}^N\sum_{j \in \mathcal{N}_i \cup\{i\}}\alpha_i \alpha_j K^f_{i,j}\\
      \text{s.t.} \quad & \Big| \sum_{j \in \mathcal{N}_i} \alpha_j \nabla K^f_h(x_i - x_j) \Big| \leq 1, \; i \in \{1, \dots, N\}, \nonumber
\end{align}
where
$w_j = \int_{\Omega} K^f_h(x - x_j) \bigl( \rho_N - \rho^* \bigr)
\diff x$ is the local mass mismatch,
$\mathcal{N}_i=\setdef{j\in V}{(i,j) \in E_{2R}(x)}$, and
$K^f_{i,j}=\int_\Omega K^f_h(x-x_i)K^f_h(x-x_j)\diff x$.  Assume, in
addition, that the agent locations become dense in $\Omega$ as
$N\to\infty$, and that the maximizers $\phi_N^*$
are $1$-Lipschitz as $N \rightarrow \infty$; \footnote{This condition is
  satisfied when the agent group become dense, and the kernel
  potentials have uniformly bounded Hessians: thus, the gradient
  constraint enforced at agent locations will extend to $\Omega$ with
  an error that vanishes as the inter-agent distance tends to zero.},
that is:
  \begin{equation}\label{eq:lip-bound}
    \sup_{\Omega}|\nabla\phi_N^*|\leq 1+\eta_N,\quad \eta_N\to 0, \quad \text{as }N\rightarrow \infty.
      \end{equation} 
      Then, the optimization problem can be solved in a distributed
      way over $\mathcal{G}_{2R}$. Furthermore, there exists a
      sequence $h=h_N\to0$ such that $\phi_N^*\to\phi^*$ uniformly as
      $N\to\infty$, where $\phi^*$ is the maximizer
      of~\eqref{eq:dual-mpc-general}.
\end{lemma}  
\begin{proof}
  To obtain the discretized problem, we substitute the density and
  potential field kernel sums, and enforce the constraint
  of~\eqref{eq:dual-mpc-general} at agent locations. With this
  formulation, $K^f_{i,j}>0$ only if the kernels of agents $i,j$
  overlap; that is, if $\dist(x_i,x_j)\leq 2R$.  Moreover, the
  gradient of the objective and the constraints are spatially
  distributed over $\mathcal{G}_{2R}$~\cite{FB-JC-SM:09} as their
  $i$-th components depend only on
  $\{(x_j,\alpha_j)\}_{j\in\mathcal {N}_i}$ and on $\rho^*$ restricted
  to $B_R(x_i)$.  As the kernel density converges to the mean-field limit,
  the discrete objective $J_N$ converges to the continuous one uniformly on the set
  of functions with bounded Lipschitz constant. The maximizers $\phi_N^*$
  of~\eqref{eq:smooth-kantorovich} exist and are unique: since the
  agent locations $x_i$ are distinct and $K^f_h$ is a strictly
  positive-definite kernel, the matrix with entries $K^f_{i,j}$ is
  strictly positive definite; thus, the objective
  $J_N(\phi_N)$ is a strongly concave quadratic function over a closed
  convex subset.  Moreover, $\phi^*_N$ are uniformly bounded (as
  $\alpha=0$ is feasible, and a solution is attained by a 
  finite $\alpha$) and asymptotically $1$-Lipschitz; hence, by the
  Arzel\`a--Ascoli theorem, $\{\phi_N^*\}$ has accumulation points
  $\hat\phi$, that are all feasible
  for~\eqref{eq:dual-mpc-general}. At the same time, $\phi^*$ is
  $1$-Lipschitz, so it can be approximated in $C^1$ by kernel
  potentials which, rescaled, are feasible
  for~\eqref{eq:smooth-kantorovich}. For each $h$ this holds for $N$
  large enough; thus, choosing a $h=h_N\to0$, we obtain feasible
  $\overline{\phi}_N\to\phi^*$ uniformly. Since
  $J_N( \overline{\phi}_N)\le J_N(\phi_N^*)$, passing to the limit
  gives $J(\phi^*)\le J(\hat\phi)$ for every accumulation point
  $\hat\phi$, and by uniqueness $\hat\phi=\phi^*$. Therefore,
  $\phi_N^*\to\phi^*$ uniformly as $N\to\infty$, with $h=h_N$.
\end{proof}
In practice, \eqref{eq:smooth-kantorovich} is solved by a distributed
primal--dual algorithm over $\mathcal{G}_{2R}$.

\subsubsection*{Low-level MPC}
At each planning time $t_\ell$, agent $i$ runs the primal--dual iteration above to
obtain $\phi_i$ and holds $\supscr{v}{nom}_i:=-\nabla\phi_i(x_i(t_\ell))$ over
$[t_\ell,t_{\ell+1})$, which defines the reference
$\supscr{x}{ref}_i(\tau)=x_i(t_\ell)+(\tau-t_\ell)\,\supscr{v}{nom}_i$. Each agent has
state $z_i=(x_i,\xi_i)$, input $u_i$, and dynamics
\begin{equation*}
  \dot z_i=f(z_i)+g(z_i)\,u_i+\big(d(x_i,t),\,0\big), \quad i \in\until{N}.
\end{equation*}
Let $t^k:=k\,\delta t$ be the control instants. As $d$ does not affect $\xi_i$ directly, it is
estimated from the residual between the measured position and the position
$\hat x_i(t^k)$ predicted by the drift-free model from $z_i(t^{k-1})$ under the input
held on $[t^{k-1},t^k)$:
\begin{equation}\label{eq:drift-est}
  \hat d_i^{\,k}=(1-\theta)\,\hat d_i^{\,k-1}+\theta\,\frac{x_i(t^k)-\hat x_i(t^k)}{\delta t},\qquad\theta\in(0,1].
\end{equation}
If $t\mapsto d(x_i(t),t)$ is $L_d$-Lipschitz and the position dynamics are $L_f$-Lipschitz in $x$, then 
 after a transient $(1-\theta)^k$,
$|d(x_i(\tau),\tau)-\hat d_i^{\,k}|\le\big((\theta^{-1}+\tfrac12)L_d+\tfrac12L_f\|d\|_\infty\big)\,\delta t+O(\delta t^2)$,
for $\tau\in[t^k,t^{k+1})$. Assuming control authority over the drift, let $(\xi^s_i(\tau),u^s_i(\tau))$ be such that
$\supscr{z}{ref}_i:=(\supscr{x}{ref}_i,\xi^s_i)$ solves
$\dot{\supscr{z}{ref}_i}=f(\supscr{z}{ref}_i)+g(\supscr{z}{ref}_i)\,u^s_i+(\hat d_i^{\,k},0)$ with
$\dot{\supscr{x}{ref}_i}=\supscr{v}{nom}_i$, and solve
{\small
\begin{equation}\label{eq:low_mpc}
\begin{aligned}
\min_{u}\ &\int_{t^k}^{t^k+T_c}\!\!\big(\|x-\supscr{x}{ref}_i\|_Q^2+\|u-u^s_i\|_R^2\big)\diff\tau+\|\zeta(t{+}T_c)\|_M^2\\
\text{s.t.}\ &\dot z=f(z)+g(z)\,u+(\hat d_i^{\,k},0),\quad z(t)=z_i(t),
\end{aligned}
\end{equation}}
with $\zeta:=z-\supscr{z}{ref}_i$, $Q,R\succ0$, horizon $T_c=\Delta t$, and terminal weight $M\succeq0$.
The system then operates at different scales; first
the top level replans every $\Delta t$: agents iterate the primal--dual algorithm, then read off
$\supscr{v}{nom}_i$ and set the reference. Within each interval, agents
re-solve~\eqref{eq:low_mpc} every $\delta t=\Delta t/n$, apply the first input,
measure $z_i$, and update $\hat d_i$ through~\eqref{eq:drift-est}. Both solves are
taken to be fast relative to $\delta t$. Macroscopically, the perturbation $v$
collects three errors: the tracking error $\supscr{e}{tr}$ of~\eqref{eq:low_mpc},
driven by the residual $d-\hat d_i$; the error $\supscr{e}{alg}$ of the distributed
algorithm with respect to the exact solution of~\eqref{eq:smooth-kantorovich}
(finitely many iterations and the held command); and the discretization error
$\supscr{e}{discr}$ of~\eqref{eq:kernel-funs}. By the triangle inequality,
\begin{equation}\label{eq:err-bound}
  \|v\|_{L^2(\rho_N)}\le \supscr{e}{tr}+\supscr{e}{alg}+\supscr{e}{discr}.
\end{equation}
\section{Bounded Stability to Bounded Mismatch}\label{sec:stability}

Here, we quantify the effect of the low-level mismatch and
of the target motion on the macroscopic closed-loop flow for a general
regularization $\mathcal{F}$, and then particularize the result to the
$L^2$ and $\kl$ penalties. To this end, we follow the distributional
input-to-state stability (dISS) approach of~\cite{GP-SM:26}, applied
here with $W_1$, with the optimal value of the regularized
problem~\eqref{eq:dual-mpc-general} as Lyapunov functional.

Assume that the conditions of Proposition~\ref{prop:fenchel-dual}
hold, and consider the continuous-time flows of the swarm density
$\rho_t$ and of the target density $\rho^*_t$:
\begin{equation}\label{eq:traj}
    \partial_t\rho=-\nabla\cdot\big(\rho\,(-\nabla\phi^*+v)\big),\qquad
    \partial_t\rho^*=-\nabla\cdot\big(\rho^*v^*\big),
\end{equation}
where $\phi^*$ is the maximizer of~\eqref{eq:dual-mpc-general} for the
pair $(\rho_t,\rho^*_t)$, $v$ is the perturbation of the nominal
velocity $-\nabla\phi^*$ as described in~\eqref{eq:err-bound}, and
$v^*$ is the vector field transporting the target distribution.
The trajectory equations are subject to no-flux boundary conditions on
$\Omega$.

Next, we consider the optimal value of~\eqref{eq:dual-mpc-general},
with $(\rho_0,\rho^*_{T_H})$ replaced by $(\rho,\rho^*)$, as a
Lyapunov functional $V(\rho)$. We assume that $\mathcal{F}\geq0$ with
$\mathcal{F}(\rho^*)=0$, for all time, and view
$\mathcal{F}(\rho) \equiv \mathcal{F}(\rho\,;\rho^*)$ to make the
dependence on the target explicit. In this way, we denote by
$\frac{\delta\mathcal{F}}{\delta\rho}|_\sigma$ and
$\frac{\delta\mathcal{F}}{\delta\rho^*}|_\sigma$ its first variations
with respect to each argument, evaluated at $(\sigma,\rho^*)$. Let
$\sigma$, the intermediate state, be the maximizer
in~\eqref{eq:fenchel_final_relation} for $\phi=\phi^*$.
By~\eqref{eq:ineq-1}, the functional can be rewritten as:
\begin{equation*}
  V(\rho)=\int_\Omega\phi^*(\diff\rho-\diff\rho^*)-\mathcal{G}(\phi^*)
  =W_1(\rho,\sigma)+\mathcal{F}(\sigma).
\end{equation*}
Note that, since the target motion is unknown to the agents, $\rho^*_{t+T_H}$
is replaced by the current $\rho^*_t$; the resulting lag is quantified
in the following.
\begin{theorem}\label{prop:stab}
  Consider the swarm and target absolutely continuous densities
  $\rho_t,\rho_t^*$ solving~\eqref{eq:traj} over a convex compact domain
  $\Omega$ with diameter $D$, let $\sigma_t$ be the intermediate state
  associated with $(\rho_t,\rho^*_t)$, and define:

  {\small
  \begin{equation*}
    R_t:=\int_\Omega\nabla\frac{\delta\mathcal{F}}{\delta\rho}\Big|_{\sigma_t}\cdot v^*\diff\sigma_t
    +\int_\Omega\nabla\frac{\delta\mathcal{F}}{\delta\rho^*}\Big|_{\sigma_t}\cdot v^*\diff\rho^*_t.
  \end{equation*}
  }
  Assume that there exist $r>0$, $\nu_{\mathcal{F}}\geq0$ and
  $\theta\in\mathcal{K}_\infty$ such that, for all $t\geq0$ and with
  $\sigma=\sigma_t$:\\
  \textbf{(A1)} $\sigma\geq 0$ a.e.;\\
  \textbf{(A2)} $\mathcal{F}(\sigma)\leq\frac{1}{r}\int_\Omega\big|\nabla\frac{\delta\mathcal{F}}{\delta\rho}|_\sigma\big|^2\diff\sigma$;\\
  \textbf{(A3)} $W_1(\sigma,\rho^*_t)\leq\theta\big(\mathcal{F}(\sigma)\big)$;\\
  \textbf{(A4)} $|R_t|\leq\nu_{\mathcal{F}}\,\mathcal{F}(\sigma)$, with $\nu_{\mathcal{F}}<r/2$.\\
  Then, the trajectory $\rho_t$ satisfies the following bound with
  respect to the moving target distribution:

  {\small
  \begin{equation*}
    W_1(\rho_t,\rho_t^*)\leq \psi^{-1}\Big(e^{-ct}W_1(\rho_0,\rho_0^*)+\frac{1}{2}\int_0^t e^{-c(t-\tau)}\Delta_\tau\diff\tau\Big),
  \end{equation*}
  }
  where $\Delta_t:=\int_{\Omega^2}|v(x)-v^*(y)|^2\diff\gamma_t(x,y)$,
  with $\gamma_t$ an optimal coupling between $\rho_t$ and $\sigma_t$. In
  addition, $c:=\kappa\big(\frac{1}{2}-\frac{\nu_{\mathcal{F}}}{r}\big)>0$,
  $\kappa:=\big(D+\frac{1}{r}\big)^{-1}$; and
  $\psi^{-1}\in\mathcal{K}_\infty$ is given by $\psi^{-1}(s):=s+\theta(s)$.
\end{theorem}

\begin{proof}
  We drop the subindex $t$ when no confusion arises. Note first that,
  following the stationarity reasoning from Lemma~\ref{le:reg-penalty},
  the intermediate state $\sigma$, well defined due to \textbf{(A1)},
  satisfies $\frac{\delta\mathcal{F}}{\delta\rho}|_\sigma=\phi^*-\lambda$,
  with $\lambda\in\real$ the multiplier of the mass constraint; that is:
  \begin{equation}\label{eq:kkt-gen}
    \nabla\frac{\delta\mathcal{F}}{\delta\rho}\Big|_\sigma=\nabla\phi^*
    \quad\text{a.e.~in }\Omega.
  \end{equation}
  In order to verify the Lyapunov behavior of the functional, we will
  first show that:
  \begin{equation}\label{eq:v-bounds-gen}
    \psi(W_1(\rho,\rho^*))\leq V(\rho) \leq W_1(\rho,\rho^*),
  \end{equation}
  where $\psi\in\mathcal{K}_\infty$. The second inequality follows from
  $\mathcal{G}\geq 0$.
  The first one is shown below.
  We define the set $M$ where the constraint is
  active, that is:
  \begin{equation*}
    M:=\setdef{x\in\Omega}{|\nabla\phi^*(x)|\geq 1}.
  \end{equation*}
  Then, one has:
  \begin{equation*}
    \int_{\Omega} |\nabla\phi^* |^2 \diff \rho\geq \int_M  \diff \rho =\rho(M).
  \end{equation*}
  On the other hand, for an optimal coupling $\gamma$ between $\rho$ and
  $\sigma$, we have:

  {\small
  \begin{equation*}
    W_1(\rho,\sigma)=\int_{\Omega^2} |x-y|\diff\gamma(x,y)
    \leq D\int_{x\neq y}\diff \gamma(x,y)\leq D\rho(M),
  \end{equation*}
}

where we used that $\phi^*$ is a Kantorovich potential between $\rho$
and $\sigma$, see Proposition~\ref{prop:fenchel-dual}, that the
optimal coupling does not transport mass where $|\nabla\phi^*|<1$,
i.e., in the region $M^c$~\cite{FS:15}, and that the first marginal of
$\gamma$ is $\rho$.  Thus:
  \begin{equation}\label{eq:w1-nabla-gen}
    \int_{\Omega} |\nabla\phi^*|^2\diff \rho\geq \frac{W_1(\rho,\sigma)}{D}.
  \end{equation}
  Moreover, since $\nabla\phi^*(x)=\nabla\phi^*(y)$ for
  $\gamma$-a.e.~$(x,y)$, as $\nabla\phi^*$ is constant along the rays
  associated with an optimal coupling~\cite[Chapter~3]{FS:15}, and the
  marginals of $\gamma$ are $\rho$ and~$\sigma$, we have:
  \begin{equation}\label{eq:rho-sigma}
  \begin{aligned}
    \int_\Omega|\nabla\phi^*|^2\diff\rho&=\int_{\Omega^2}|\nabla\phi^*(x)|^2\diff\gamma(x,y)\\
    &=\int_{\Omega^2}|\nabla\phi^*(y)|^2\diff\gamma(x,y)=\int_\Omega|\nabla\phi^*|^2\diff\sigma.
  \end{aligned}
  \end{equation}
  Hence, \textbf{(A2)} and~\eqref{eq:kkt-gen} yield:
  \begin{equation}\label{eq:f-nabla-gen}
    \int_\Omega|\nabla\phi^*|^2\diff\rho\geq r\,\mathcal{F}(\sigma).
  \end{equation}
  Putting~\eqref{eq:w1-nabla-gen} and~\eqref{eq:f-nabla-gen} together in
  $V(\rho)=W_1(\rho,\sigma)+\mathcal{F}(\sigma)$, we obtain:
  \begin{equation}\label{eq:v-nabla-gen}
    \int_{\Omega}|\nabla\phi^*|^2\diff\rho\geq\kappa V(\rho),
    \quad  \kappa:=\Big(D+\frac{1}{r}\Big)^{-1}.
  \end{equation}
  The decomposition also provides an explicit
  $\psi\in\mathcal{K}_\infty$ for the lower bound
  in~\eqref{eq:v-bounds-gen}. Indeed, as both terms in the
  decomposition are nonnegative, $W_1(\rho,\sigma)\leq V(\rho)$ and
  $\mathcal{F}(\sigma)\leq V(\rho)$. Then, by the triangle inequality,
  \textbf{(A3)}, and the monotonicity of $\theta$:

  {\small
  \begin{equation*}
  \begin{aligned}
  W_1(\rho,\rho^*)
  &\leq W_1(\rho,\sigma)+W_1(\sigma,\rho^*) \\
  &\leq V(\rho)+\theta\big(\mathcal{F}(\sigma)\big)
  \leq V(\rho)+\theta\big(V(\rho)\big).
  \end{aligned}
  \end{equation*}
  }
  Consequently, $W_1(\rho,\rho^*)\leq \psi^{-1}(V(\rho))$, with
  $\psi^{-1}(s):=s+\theta(s)$. Its inverse
  $\psi\in\mathcal{K}_\infty$ gives
  $\psi\big(W_1(\rho,\rho^*)\big)\leq V(\rho)$.

  After this, we bound the decay along the trajectory flow:
  \begin{equation*}
    \frac{d}{dt} V(\rho) = \int_{\Omega}
    \frac{\delta V}{\delta \rho} \partial_t \rho \, \diff x
    + \int_{\Omega} \frac{\delta V}{\delta \rho^*} \partial_t \rho^* \, \diff x.
  \end{equation*}
  In order to compute the first variation, we apply Danskin's
  theorem~\cite{JFB-AS:13} to differentiate through the
  optimization. The feasible set does not depend on $(\rho,\rho^*)$,
  and the maximizer $\phi^*$ is unique up to an additive constant.
  Hence, the first variation of $V$ reduces to the partial derivative of
  the objective function evaluated at $\phi^*$; that is,
  $\frac{\delta V}{\delta\rho}=\phi^*$ and
  $\frac{\delta V}{\delta\rho^*}=-\phi^*-\frac{\delta\mathcal{G}}{\delta\rho^*}|_{\phi^*}$.
  Applying Danskin's theorem again, we get
  $\frac{\delta\mathcal{G}}{\delta\rho^*}|_{\phi^*}=-\phi^*-\frac{\delta\mathcal{F}}{\delta\rho^*}|_\sigma$, hence:

  {\small
  \begin{equation*}
    \frac{d}{dt} V(\rho) = \int_{\Omega} \phi^* \partial_t \rho \, \diff x
    + \int_{\Omega} \frac{\delta\mathcal{F}}{\delta\rho^*}\Big|_\sigma\, \partial_t \rho^* \, \diff x.
  \end{equation*}
  }
  Substituting the trajectories in~\eqref{eq:traj} we obtain:

  {\small
  \begin{equation*}
    \frac{d}{dt} V(\rho) = -\int_{\Omega} \phi^*
    \nabla\cdot\big(\rho (-\nabla\phi^*+v)\big) \diff x
    - \int_{\Omega}  \frac{\delta\mathcal{F}}{\delta\rho^*}\Big|_\sigma \nabla\cdot(\rho^* v^*) \diff x.
  \end{equation*}
  } Under no-flux boundary conditions, we apply integration by parts
  to obtain:

  {\small
  \begin{equation*}
    \frac{d}{dt} V(\rho) = -\int_{\Omega} |\nabla\phi^*|^2
    \diff \rho +\int_{\Omega}\nabla \phi^*\cdot v\diff \rho
   +\int_\Omega \nabla\frac{\delta\mathcal{F}}{\delta\rho^*}\Big|_\sigma \cdot v^* \diff\rho^*,
  \end{equation*}
} where we have used $\rho\diff x=\diff \rho$ due to absolute
continuity. By the definition of $R_t$ and~\eqref{eq:kkt-gen}, the
last term equals $R_t-\int_\Omega\nabla\phi^*\cdot v^*\diff\sigma$, so
the last two terms in the previous equality can be rewritten as:

  {\small
  \begin{equation*}
    \int_{\Omega^2}\big(\nabla \phi^*(x)\cdot v(x)-\nabla\phi^*(y)
    \cdot v^*(y)\big)\diff \gamma(x,y) +R_t.
  \end{equation*}
  }
  To handle the first term, we use again that
  $\nabla\phi^*(x)=\nabla\phi^*(y)$ for $\gamma$-a.e.~$(x,y)$, and apply
  Young's inequality:

  {\small
  \begin{equation*}
    \int_{\Omega^2} \nabla\phi^*(x) \cdot \big(v(x)-v^*(y)\big)\diff \gamma_t(x,y)
    \leq \frac{1}{2} \int_\Omega |\nabla\phi^*|^2\diff \rho+\frac{1}{2}\Delta_t,
  \end{equation*}
  } where we have marked the dependence of $\gamma$ w.r.t.~$t$.
  For the term $R_t$, \textbf{(A4)} and~\eqref{eq:f-nabla-gen} give:
  \begin{equation*}
    |R_t|\leq\nu_{\mathcal{F}}\,\mathcal{F}(\sigma)\leq
    \frac{\nu_{\mathcal{F}}}{r}\int_\Omega|\nabla\phi^*|^2\diff\rho.
  \end{equation*}
  Now, we put everything together to obtain the bound:
  \begin{equation*}
    \frac{d}{d t}V(\rho)\leq -
    \Big(\frac{1}{2}-\frac{\nu_{\mathcal{F}}}{r}\Big)
    \int_{\Omega}|\nabla\phi^*|^2\diff\rho+\frac{1}{2}\Delta_t,
  \end{equation*}
  which, by~\eqref{eq:v-nabla-gen}, finally yields:
  \begin{equation*}
    \frac{d}{d t}V(\rho)\leq -c V(\rho)
    +\frac{1}{2}\Delta_t,
  \end{equation*}
  where $c=\kappa\big(\frac{1}{2}-\frac{\nu_{\mathcal{F}}}{r}\big)>0$
  by \textbf{(A4)}. Applying Gr\"onwall's inequality, we obtain:
  \begin{equation*}
    V(\rho_t)\le e^{-ct}V(\rho_0)+\frac{1}{2}\int_{0}^{t}e^{-c(t-\tau)}\Delta_\tau\,\diff\tau.
  \end{equation*}
  Finally, using both bounds in~\eqref{eq:v-bounds-gen}, together with
  the monotonicity of $\psi^{-1}$, yields the claimed result.
\end{proof}
Next, we find conditions under which the assumptions of
Theorem~\ref{prop:stab} hold for the $L^2$ and $\kl$ penalties.
\begin{lemma}\label{le:assump-l2}
  Assume that, $\forall t\geq0$, $\rho^*_t$ is $\epsilon$-Lipschitz,
  the swarm density is strictly positive, $\rho_t\geq m>0$, and the
  target velocity field satisfies $\|\nabla\cdot v^*\|_\infty\leq\nu$,
  with $\nu\epsilon D^2<\pi^2 m$. Then, the assumptions of
  Theorem~\ref{prop:stab} hold for the $L^2$ penalty
  $\mathcal{F}(\sigma)=\frac{1}{2\epsilon}\|\sigma-\rho^*\|_{L^2}^2$, with
  $r=\frac{2\pi^2m}{\epsilon D^2}$, $\theta(s)=D\sqrt{\epsilon|\Omega|s/2}$,
  and $\nu_{\mathcal{F}}=\nu$.
\end{lemma}
\begin{proof}
  Lemma~\ref{le:positivity} guarantees \textbf{(A1)} under the
  $\epsilon$-Lipschitz condition. For the remaining assumptions, note
  that
  $\frac{\delta\mathcal{F}}{\delta\rho}|_\sigma=\frac{1}{\epsilon}(\sigma-\rho^*)=\phi^*$
  and $\frac{\delta\mathcal{F}}{\delta\rho^*}|_\sigma=-\phi^*$, so that
  $\mathcal{F}(\sigma)=\frac{\epsilon}{2}\int_\Omega(\phi^*)^2\diff x$.
  Moreover, $\phi^*$ has zero mean, since $\epsilon\phi^*=\sigma-\rho^*$
  with $\sigma,\rho^*\in\P$.

  For \textbf{(A2)}, we use the Poincar\'e-Wirtinger
  inequality~\cite{LCE:98}, with the constant of~\cite{LEP-HFW:60} for
  convex domains, which applies to $\phi^*\in W^{1,\infty}(\Omega)$ as it
  has zero mean. Together with~\eqref{eq:rho-sigma} and $\rho\geq m$, it
  gives:

  {\small
  \begin{equation*}
  \begin{aligned}
    \int_{\Omega}|\nabla\phi^*|^2\diff\sigma
    =\int_{\Omega}|\nabla\phi^*|^2\diff\rho \geq
    \\  m\int_{\Omega}|\nabla\phi^*|^2\diff x
    \geq \frac{m\pi^2}{D^2}\int_{\Omega}|\phi^*|^2\diff x
    =\frac{2\pi^2m}{\epsilon D^2}\,\mathcal{F}(\sigma).
   \end{aligned}
  \end{equation*}
  }
  As $\nabla\frac{\delta\mathcal{F}}{\delta\rho}|_\sigma=\nabla\phi^*$, this
  is \textbf{(A2)} with $r=\frac{2\pi^2m}{\epsilon D^2}$.

  For \textbf{(A3)}, we have
  $W_1(\sigma,\rho^*)\leq\frac{D}{2}\|\sigma-\rho^*\|_{L^1}$, from the dual
  formulation~\eqref{eq:dual-rn}. Then, as
  $\|\sigma-\rho^*\|_{L^1}\leq\sqrt{|\Omega|}\|\sigma-\rho^*\|_{L^2}$:
  \begin{equation*}
    W_1(\sigma,\rho^*)\leq\frac{D}{2}\sqrt{|\Omega|}\,\|\sigma-\rho^*\|_{L^2}
    =D\sqrt{\frac{\epsilon|\Omega|}{2}\,\mathcal{F}(\sigma)}.
  \end{equation*}
  Finally, for \textbf{(A4)}, by the definition of $R_t$ and the first
  variations above:
  \begin{equation*}
   R_t=\int_\Omega\nabla\phi^*\cdot v^*\diff\sigma
    -\int_\Omega\nabla\phi^*\cdot v^*\diff\rho^*
    =\int_\Omega\nabla\phi^*\cdot v^*\,(\sigma-\rho^*)\diff x.
  \end{equation*}
  Then, we substitute $\sigma-\rho^*=\epsilon\phi^*$, apply the chain
  rule and integration by parts to obtain:

  {\small
  \begin{equation*}
    \int_\Omega\nabla \phi^*\cdot v^*\,(\epsilon\phi^*)\diff x
    =\int_\Omega  v^*\cdot\frac{\epsilon}{2}\nabla(\phi^*)^{2}\diff x
    =-\int_\Omega\frac{\epsilon}{2}(\phi^*)^{2}\nabla\cdot v^* \diff x,
  \end{equation*}
  }
  where the boundary term vanishes due to the no-flux condition. Given
  the bound $\|\nabla\cdot v^*\|_\infty\leq \nu$, we can write:
  \begin{equation*}
    |R_t|\leq\nu\,\frac{\epsilon}{2}\int_\Omega(\phi^*)^2\diff x
    =\nu\,\mathcal{F}(\sigma).
  \end{equation*}
  That is, \textbf{(A4)} holds with $\nu_{\mathcal{F}}=\nu$, and
  $\nu_{\mathcal{F}}<r/2$ is equivalent to $\nu\epsilon D^2<\pi^2m$.
\end{proof}
\begin{lemma}\label{le:assump-kl}
  Assume that, $\forall t\geq0$, $\rho^*_t>0$ a.e.~and $\rho^*_t$
  satisfies a logarithmic Sobolev inequality (LSI)~\cite{CV:08} with
  constant $\alpha>0$. Then, the assumptions of
  Theorem~\ref{prop:stab} hold for the $\kl$ penalty
  $\mathcal{F}(\sigma)=\frac{1}{\epsilon}\kl(\sigma\,\|\,\rho^*)$,
  with $r=\frac{2\alpha}{\epsilon}$, $\theta(s)=D\sqrt{\epsilon s/2}$,
  and $\nu_{\mathcal{F}}=0$, for any velocity field $v^*$.
\end{lemma}
\begin{proof}
  By Lemma~\ref{le:reg-penalty}, $\sigma=e^{\epsilon\phi^*}\rho^*\in\P$,
  so \textbf{(A1)} holds as $\rho^*>0$ a.e. For the remaining
  assumptions, note that
  $\frac{\delta\mathcal{F}}{\delta\rho}|_\sigma
  =\frac{1}{\epsilon}\big(\log\frac{\sigma}{\rho^*}+1\big)
  =\phi^*+\frac{1}{\epsilon}$ and
  $\frac{\delta\mathcal{F}}{\delta\rho^*}|_\sigma
  =-\frac{1}{\epsilon}e^{\epsilon\phi^*}$.

  For \textbf{(A2)}, we apply the LSI to $\sigma$. Since
  $\log(\sigma/\rho^*)=\epsilon\phi^*$, we obtain:

  {\small
  \begin{equation*}
    \epsilon\mathcal{F}(\sigma)=\kl(\sigma\,\|\,\rho^*)
    \leq\frac{\epsilon^2}{2\alpha}\int_\Omega|\nabla\frac{\delta\mathcal{F}}{\delta\rho}|_\sigma|^2\diff\sigma=\frac{\epsilon^2}{2\alpha}\int_\Omega|\nabla\phi^*|^2\diff\sigma.
  \end{equation*}
  }
This is \textbf{(A2)} with $r=\frac{2\alpha}{\epsilon}$.

  For \textbf{(A3)}, we have
  $W_1(\sigma,\rho^*)\leq\frac{D}{2}\|\sigma-\rho^*\|_{L^1}$, as in
  Lemma~\ref{le:assump-l2}, and, by Pinsker's
  inequality~\cite{ABT:09}:

  {\small
  \begin{equation*}
    W_1(\sigma,\rho^*)\leq\frac{D}{2}\sqrt{2\kl(\sigma\,\|\,\rho^*)}
    =D\sqrt{\frac{\epsilon}{2}\,\mathcal{F}(\sigma)}.
  \end{equation*}
  }

  Finally, for \textbf{(A4)}, by the definition of $R_t$, the first
  variations above, 
  $\nabla e^{\epsilon\phi^*}=\epsilon e^{\epsilon\phi^*}\nabla\phi^*$,
  and $e^{\epsilon\phi^*}\rho^*\diff x=\diff\sigma$:

  {\small
  \begin{equation*}
    R_t=\int_\Omega\nabla\phi^*\cdot v^*\diff\sigma
    -\int_\Omega\nabla\phi^*\cdot v^*\diff\sigma=0.
  \end{equation*}
  }
  That is, \textbf{(A4)} holds with $\nu_{\mathcal{F}}=0$, without any
  condition on $v^*$. This reflects the invariance of $\kl$ under a
  common pushforward of $\sigma$ and $\rho^*$.
\end{proof}
Plugging these constants into Theorem~\ref{prop:stab}, we obtain, for
the $L^2$ penalty,
$\kappa=\big(D(1+\frac{\epsilon D}{2\pi^2m})\big)^{-1}$,
$c=\kappa\big(\frac12-\frac{\nu\epsilon D^2}{2\pi^2m}\big)$ and
$\psi^{-1}(s)=s+D\sqrt{\epsilon|\Omega|s/2}$; and, for the $\kl$ penalty,
$\kappa=\big(D+\frac{\epsilon}{2\alpha}\big)^{-1}$,
$c=\kappa/2$ and $\psi^{-1}(s)=s+D\sqrt{\epsilon s/2}$.
Regarding the assumptions, note that satisfying LSI for a target measure
can be achieved by the Bakry--\'Emery criterion: if
$\rho^*_t\propto e^{-f_t}$ on $\Omega$ with $\nabla^2f_t\succeq\beta \Id$, $\beta>0$,
then LSI holds; alternatively, any density bounded from below and above,
$0<m_1\leq\rho^*_t\leq m_2$, also satisfies it, see~\cite{DB-IG-ML:14}.

\begin{remark}[Exact tracking]\label{rem:v1v2}
  The term $\Delta_t$ vanishes when $v(x)=v^*(y)$ for $\gamma_t$-a.e.\
  $(x,y)$, i.e., when the perturbation reproduces the motion of the
  target along the coupling $\gamma_t$.  However, in general that does
  not happen; here $v$ collects the errors of the low
  level~\eqref{eq:err-bound}, none of which uses information on $v^*$,
  so the two fields are uncorrelated.  In particular, even without
  perturbations ($v=0$),
  $\limsup_{t\to\infty}W_1(\rho_t,\rho_t^*)\leq\psi^{-1}\big(\|v^*\|_\infty^2/(2c)\big)$,
  i.e., the swarm lags the moving target.  In future work, the motion
  of the target will be accounted for in order to reduce that
  error.\oprocend
\end{remark}

\section{Numerical Implementation}
We consider $N$ agents with states $(x_i,v_i)\in\mathbb{R}^2\times\mathbb{R}^2$ and input $u_i\in\mathbb{R}^2$, with dynamics
\footnote{The code and data used to generate the numerical results as well as videos of the tracking are available at \url{https://github.com/GuillemPascualAstivill/Swarm-Target-Tracking-with-MPC}.}
$ \dot{x}_i= v_i + d(x_i,t);\quad\dot{v}_i= u_i$.
The drift $d$
is a Taylor-Green vortex~\cite{GIT-AEG:37} field of amplitude $U$.
For the density approximation we use the following Wendland compactly
supported kernel~\cite{HW:95}:
\begin{equation}\label{eq:wendland}
K_h(x) =
\begin{cases}
\displaystyle
\frac{7}{\pi h^2}
\left(1-\frac{|x|}{h}\right)^4
\left(4\frac{|x|}{h}+1\right),
& |x| < h, \\[1.2ex]
0,
& | x | \ge h.
\end{cases}
\end{equation}
The nominal velocity
$\supscr{v_i}{nom}=-\nabla \phi(t,x_i(t))$ is computed using a
primal-dual algorithm that
solves~\eqref{eq:smooth-kantorovich}.

\begin{figure}[t]
  \centering
  \includegraphics[width=\columnwidth]{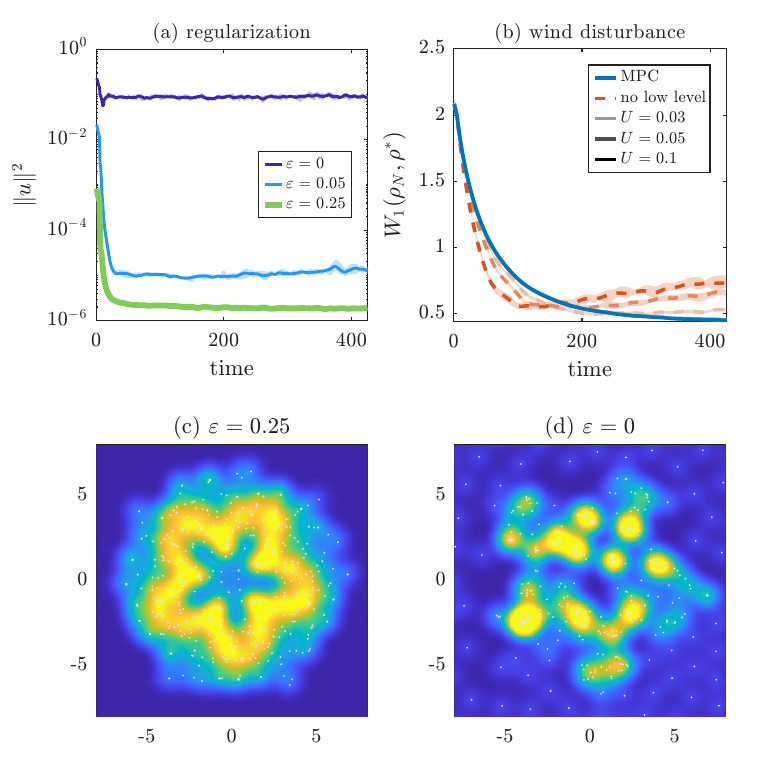}
  \caption{Simulation results for $N=500$ agents and a rotating star
    target; curves are means over 6 initial conditions (bands: range;
    10-step moving average). (a) Control effort $\|u\|^2$ per agent
    for $\varepsilon\in\{0,0.05,0.25\}$ (wind $U=0.01$). (b) $W_1$
    density error with (blue, solid) and without (orange, dashed) the
    low-level MPC, for $U\in\{0.03,0.05,0.1\}$ (light to
    dark). (c)--(d) Final agent density with ($\varepsilon=0.25$) and
    without ($\varepsilon=0$) regularization ($U=0.01$); color scale
    clipped at the peak of $\rho^*$.}
  \label{fig:sim}
\end{figure}

 The drift-free prediction in~\eqref{eq:drift-est} is $\hat x_i=x_i+\delta t\,v_i+\tfrac12\delta t^2u_i$ for the held input,
  and the constant pair is $\supscr{\xi}{s}_i=\supscr{v}{nom}_i-\hat d_i$, $\supscr{u}{s}_i=0$, which exists for every drift,
  so the estimate simply shifts the velocity reference: $\supscr{z}{ref}_i=(\supscr{x}{ref}_i,\supscr{v}{nom}_i-\hat d_i)$. 
The error $\zeta_i=z_i-\supscr{z}{ref}_i$ satisfies $\dot\zeta_i=A\zeta_i+Bu_i+E(d-\hat d_i)$, with $A=\left[\begin{smallmatrix}0&I\\0&0\end{smallmatrix}\right]$,
$B=\left[\begin{smallmatrix}0\\I\end{smallmatrix}\right]$
 and $E=\left[\begin{smallmatrix}I\\0\end{smallmatrix}\right]$.
We take $M=P$, the solution of the discrete-time Riccati equation of this model
discretized with step $\delta t$; without input constraints,~\eqref{eq:low_mpc}
then reduces to the LQR feedback.
We consider a nonconvex,
star-shaped target density that rotates rigidly; this motion is
divergence-free, so $\nu_{\mathcal{F}}=0$ in Theorem~\ref{prop:stab}. 
 We use $N=500$ agents, 
 demonstrating that the proposed method can be applied to relatively large swarms with practical computational cost.
Fig.~\ref{fig:sim} shows that the method tracks the moving target and
rejects the wind. With the low-level MPC, the density error settles at
the same low level for all $U$; this residual includes the tracking
lag of Remark~\ref{rem:v1v2}. Agents that only follow the nominal
velocity, without the low level, settle at an error that grows with
$U$, and some even leave the domain. The regularization shapes the
swarm rather than its distance to the target: without it
($\varepsilon=0$), the commanded speeds are significantly larger, the
control effort grows by four orders of magnitude
(Fig.~\ref{fig:sim}a), and the agents collapse into dense clusters
(Fig.~\ref{fig:sim}d).

\section{Conclusions}
In this paper we presented a multiscale controller for swarm
control. By leveraging optimal transport tools, we were able to relate
a terminal cost in its dynamic formulation with a regularization in
the static problem that yields strict concavity and lends itself to a
distributed implementation, as well as handling moving targets
smoothly.  At the same time, the low-level MPC is able to handle
external disturbances and control-affine dynamics.  The mismatch
between both levels is analyzed via distributional input-to-state stability of the
probability flow, allowing for a quantification of the error induced
in the Wasserstein metric. Further work will focus on incorporating
safety constraints, such as obstacle avoidance and density congestion,
improving the tracking of moving targets, and investigating
alternative regularization schemes.

\section*{Acknowledgements}
  \textit{Gemini Pro
  was used to assist with 
  code troubleshooting, checking if particular results exist in the
  literature, and for literature search. All results and references
  have been verified by the authors.}  
\bibliographystyle{unsrt} \bibliography{alias,SMD-add,SM,JC}
\end{document}